# Testing Kak's Conjecture on Binary Reciprocal of Primes and Cryptographic Applications

Sumanth Kumar Reddy Gangasani
Oklahoma State University, Stillwater

*Abstract:* This note considers reciprocal of primes in binary representation and shows that the conjecture that 0s exceed 1s in most cases continues to hold for primes less than one million. The conjecture has also been tested for ternary representation with similar results. Some applications of this result to cryptography are discussed.

## Introduction

In a note nearly 25 years ago, Kak conjectured [1] that binary reciprocals of primes have an excess of zeroes. This result is surprising because the expectation is that if a large number of reciprocals of primes are tested, the frequencies of 0s and 1s should be about the same. This conjecture was based on testing primes less that 5,471. In this note we consider Kak's conjecture for primes up to one million, showing that it continues to hold.

Assuming that this property holds in general, this will have cryptographic applications. If, say, for only 3 percent of the prime reciprocals the number of 1s exceeds the number of 0s, then a randomly chosen prime will have a 3 percent chance of satisfying this property. By combining several such choices, one will be able to generate events that have a given probability and such events can be useful in cryptographic protocols for e-commerce and other applications.

## Reciprocals of primes

The expansion of the reciprocal of prime is a decimal sequence [2]. For representation in an arbitrary base, the term d-sequence is used. Several cryptographic and communications properties of these sequences are known [3-9]. For maximum-length binary d-sequences, which are expansions of $1/p$, p prime, for which 2 is a primitive root of p, there are the same number of 0s and 1s, and such d-sequences have good randomness properties. But for other binary d-sequences the number of 0s and 1s need not be equal.

In [1] it was shown that for the first 722 primes there are only 14 anomalous ones for which the number of 1s exceeds that of 0s, which is less than 2 percent. This anomaly was found to hold for prime reciprocals in base 3 also. The binary reciprocal of prime is generated by means of the algorithm [5]:

$$a(i) = 2^i \bmod p \bmod 2$$

where *p* is the prime number. There is a similar formula to generate reciprocals of primes in other bases.

## Results of Our Experiments

The main result of our experiment for primes from 7 to 999,983 is:

**Table 1:** Proportion of 0s and 1s for primes less one million

| Cases where 0s exceed 1s | 19888 |
|---|---|
| Cases where 1s exceed 0s | 3059 |
| Equal number of 0s and 1s | 55544 |

The proportion of cases where 1s exceed 0s is 3.9 percent whereas the number of cases where 0s exceed 1s is 25.36 percent.

For better visualization, we now give the results on the difference between 0s and 1s for primes less than 65,535. As is clear from Figure 1, the cases where 0s exceed 1s are much more numerous than those where 1s exceed 0s. In Figure 2, we have only considered those cases where 1s exceed 0s and to highlight this case, the scale for the y-axis is to a smaller unit.

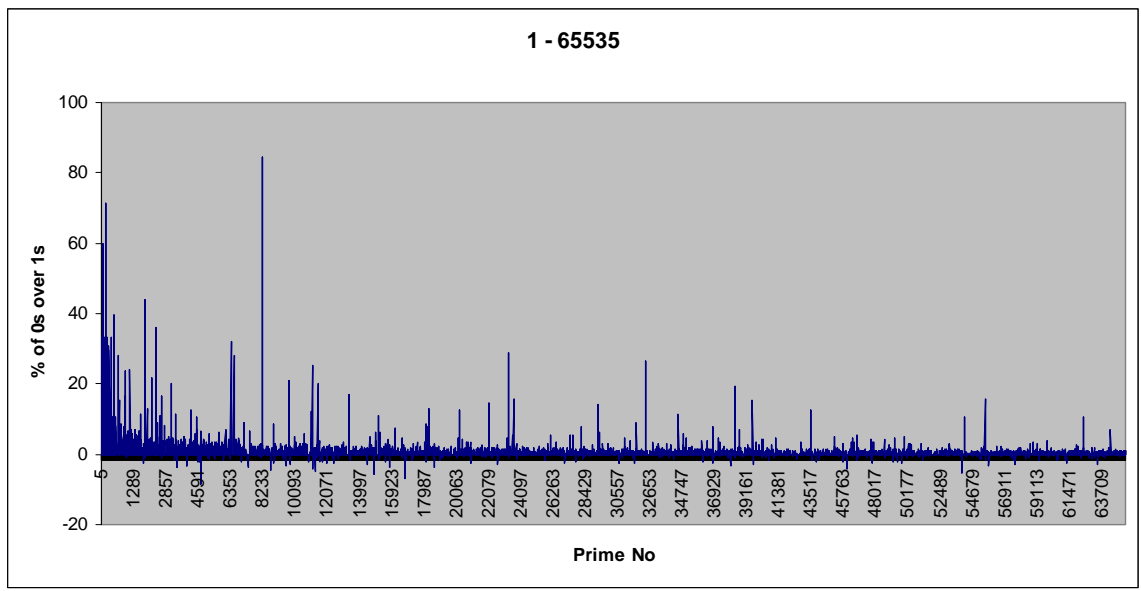

**Figure 1:** Plot of percentage difference of 0s and 1s in d-sequences where number of 0s and 1s are not equal (positive y-axis, 0s exceed 1s; negative y-axis, 1s exceed 0s).

On an average 46.71% of non maximum length sequences have different number of 0s and 1s when studied till primes less than 65,535 and 29.26% of all the sequences (including maximum length) have different 0s and 1s.

It can be clearly seen that for Mersenne prime 8191 the number of 0s is far more than the number of 1s in the d-sequence.

Figure 2 presents only those cases where the number of 1s exceeds 0s. Comparing Figures 1 and 2 it is clear that the percentage difference for such cases is much smaller than for the cases where 0s exceed 1s.

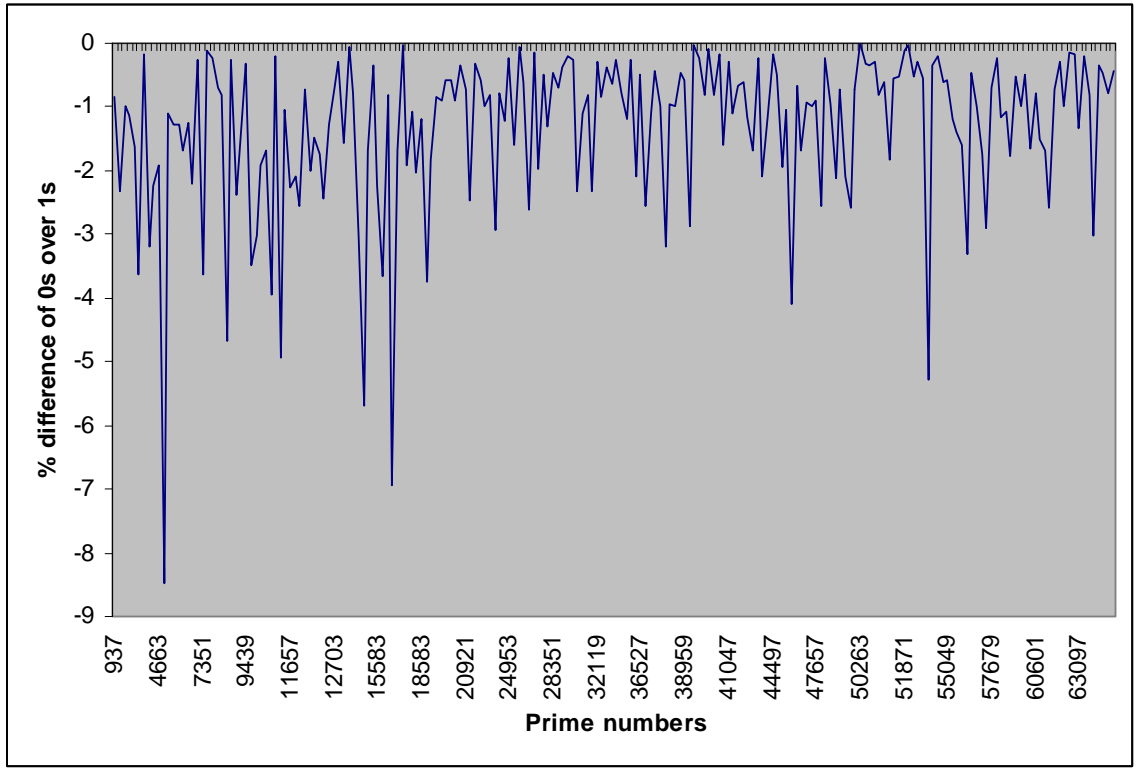

**Figure 2:** Plot of percentage difference of 0s and 1s in d-sequences where number of 0s is less than number of 1s.

It is seen that in around 3.15% of all the cases 1s are greater than 0s. In other cases, 0s are either equal to 1s or exceed 1s. Table 2 summarizes the discrepancy between 0s and 1s in the binary d-sequences in various ranges.

**Table 2**: Proportion of 0s and 1s in various ranges

| Prime No. range | 0s are greater | 1s are greater |
|---|---|---|
| 1-10000 | 320 | 30 |
| 10001-20000 | 256 | 40 |
| 20001-30000 | 265 | 24 |
| 30001-40000 | 251 | 27 |
| 40001-50000 | 240 | 31 |
| 50001-60000 | 245 | 35 |
| 60001-65535 | 131 | 19 |

## Ternary sequences

If one were to extend the original conjecture to ternary (base-3) sequences, one would expect the number of 0s to exceed the number of 1s and 2s. This is seen very clearly in Figures 3 and 4.

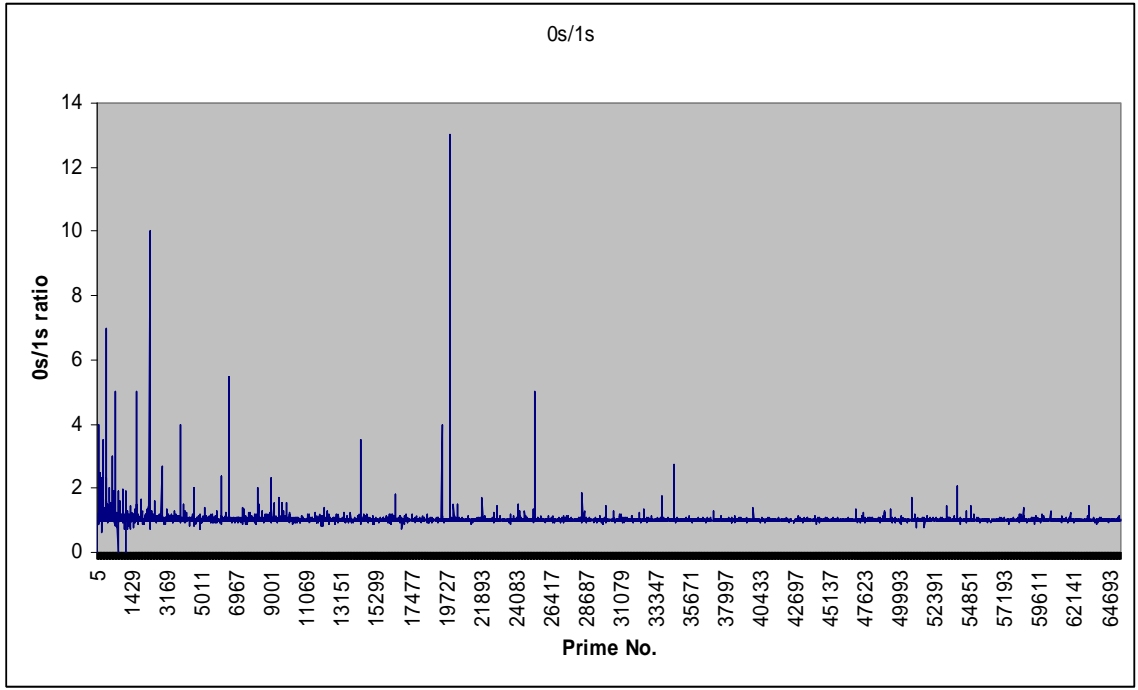

**Figure 3:** The ratio of 0s over 1s for prime reciprocals to base 3

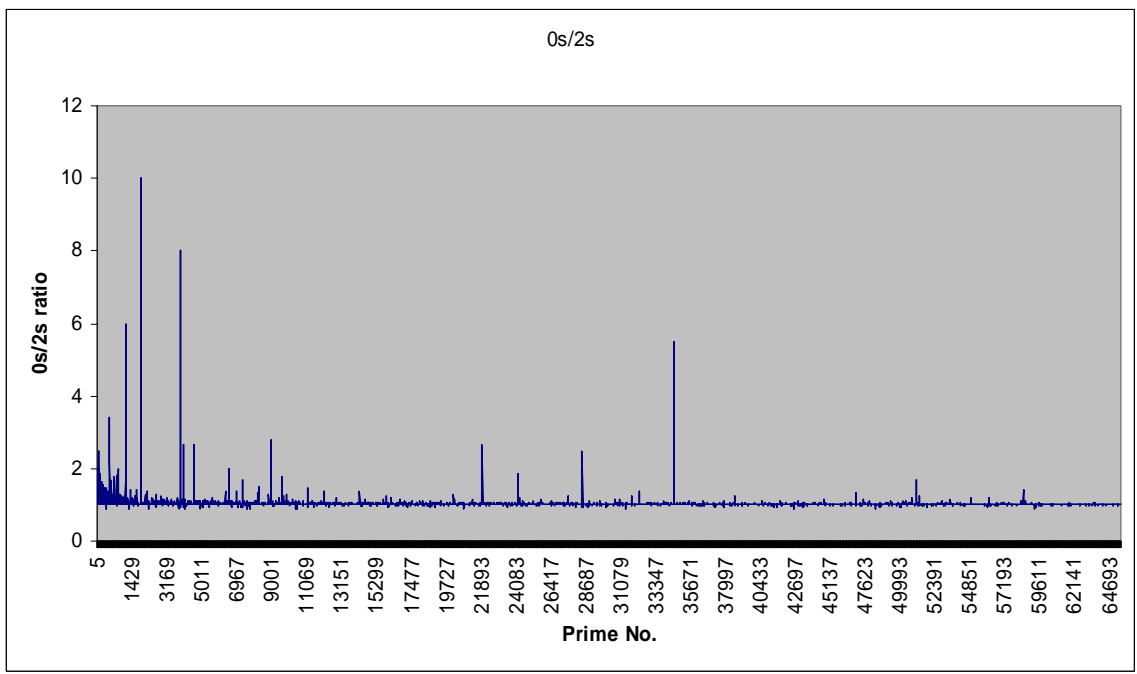

**Figure 4:** The ratio of 0s over 2s for prime reciprocals to base 3

The 0s are more numerous than 1s and 2s for a large number of cases, although as in the binary case there are situations where the reverse is true.

## Conclusions

We have shown that the conjecture that 0s exceed 1s more often than the other way round for binary reciprocals of primes continues to hold for primes up to one million. The number of cases where 0s exceeded 1s was 19,888, whereas the cases where 1s exceeded 0s was only 3,059. Since this is rather surprising, it would be worthwhile to check if this conjecture holds for still greater primes. If a proof of this conjecture can be furnished that would be of much theoretical interest to number theorists and to cryptographers.

Even if the conjecture does not hold for all primes but is true for primes in a certain range, one can take advantage of the result by requiring the parties in the cryptographic protocol to use primes in that range.